\documentclass[prd,amsfonts,amssymb,
amsmath,secnumarabic,preprint,showpacs
]{revtex4}

\usepackage{bm}
\usepackage{hyperref}

\newcommand{\ffrac}[2]{\raisebox{.5pt}%
  {\footnotesize$\displaystyle\frac{#1}{#2}$}\kern1pt}
\renewcommand{\d}{\partial}
\newcommand{\half}{\mathchoice{%
    \ffrac{1}{2}}{\frac{1}{2}}{\frac{1}{2}}{\frac{1}{2}}}
\def\napprox{\approx\hspace{-0.90em}\slash\hspace{0.38em} }

\def\cA{\mathcal{A}}

\def\cE{\mathcal{E}}

\def\cJ{\mathcal{J}}

\def\cL{\mathcal{L}}

\def\cP{\mathcal{P}}
\def\cQ{\mathcal{Q}}

\def\Q#1#2{\frac{\partial #1}{\partial #2}}

\def\varQ#1#2{\frac{\delta #1}{\delta #2}}

\begin{document}

\preprint{ULB-TH/07-10}

\title{Note on the First Law with $p$-form potentials}

\author{Geoffrey Comp\`ere}
\affiliation{%
Physique Th\'eorique et Math\'ematique, \\ Universit\'e Libre de
    Bruxelles\\and\\ International Solvay Institutes\\ Campus
    Plaine C.P. 231, B-1050 Bruxelles, Belgium
}

\begin{abstract}
The conserved charges for $p$-form gauge fields coupled to gravity
are defined using Lagrangian methods. Our expression for the
surface charges is compared with an earlier expression derived
using covariant phase space methods. Additional properties of the
surfaces charges are discussed. The proof of the first law for
gauge fields that are regular when pulled-back on the future
horizon is detailed and is shown to be valid on the bifurcation
surface as well. The formalism is applied to black rings with
dipole charges and is also used to provide a definition of energy
in plane wave backgrounds.

\end{abstract}


\maketitle

Remarkably, the first law of black hole mechanics has been
demonstrated for arbitrary perturbations around a stationary black
hole with bifurcation Killing horizon in any diffeomorphism
invariant theory of gravity~\cite{Iyer:1994ys}. Also, this law has
been shown to hold when gravity is coupled to Maxwell or
Yang-Mills fields as a consequence of conservation laws and of
geometric properties of the
horizon~\cite{Sudarsky:1992ty,Gao:2003ys}.

Recently, black rings with gauge charge along the ring, the
so-called dipole charge, have been found in five dimensional
supergravity~\cite{Emparan:2004wy}. As shown
in~\cite{Copsey:2005se}, the black ring solutions with dipole
charge have a potential which diverges at the bifurcation surface.
This implies that the computations
of~\cite{Sudarsky:1992ty,Iyer:1994ys} are not directly applicable
to that case.

Hamiltonian methods were applied to gravity coupled to a $p$-form
and a scalar field in order to explain the occurrence of dipole
charges in the first law~\cite{Copsey:2005se}. Quasilocal
formalism~\cite{Astefanesei:2005ad} as well as covariant phase
space methods~\cite{Rogatko:2005aj,Rogatko:2006xh} have also been
developed. The first aim of this paper is to improve the covariant
analysis~\cite{Rogatko:2005aj,Rogatko:2006xh} by deriving an
expression for the conserved charges taking better care of the
form factors. Following the Lagrangian methods based on
cohomological results~\cite{Barnich:2001jy,Barnich:2003xg}, our
expression for the surface charges will moreover get round the
usual ambiguities of covariant phase methods. Several properties
of these surface charges will be discussed.

It was observed in~\cite{Gao:2003ys,Rogatko:2006xh} that a
consistent thermodynamics can be done on the future event horizon
with gauge potentials that may be irregular on the bifurcation
surface if, nevertheless, the potential is regular when
pulled-back on the future horizon. We will extend the analysis
of~\cite{Rogatko:2005aj,Rogatko:2006xh} by detailing how this
regularity hypothesis allows for proving the first law in that
context. We point out that the proof of the first law is valid on
the bifurcation surface as well. We will then show that the
potential for the black rings~\cite{Emparan:2004wy} admits a
regular pull-back on the future event horizon and can thus be
treated by this method. Note that this analysis covers only
electric-type charges and not magnetic charges where the potential
is necessarily singular on the future event horizon.

Conservations laws have been defined in asymptotically flat and
anti-de Sitter backgrounds, see e.g. the seminal
works~\cite{Arnowitt:1962aa,Regge:1974zd,Abbott:1982ff}. A natural
question, raised
in~\cite{Gimon:2003xk,Hubeny:2003ug,Hashimoto:2004ve}, is how mass
can be defined in asymptotic plane wave geometries. We show in the
last section that the conserved charges defined in this paper can
be used in this context and lead to the correct first law.

In what follows, we will consider the action
\begin{eqnarray}
 S[g,\mathbf A,\phi] &=&  \frac{1}{16 \pi G}\int  \left[\star 1\, R -
\star 1\, \frac{1}{2}\d_\mu\chi \d^\mu \chi - \frac{1}{2}
e^{-\alpha \chi} \mathbf H \wedge \star \mathbf H
\right],\label{act1}
\end{eqnarray}
where $\chi$ is a dilaton and $\mathbf H = d \mathbf A$ is the
field strength of a $p$-form $\mathbf A$, $p\geq 1
$~\footnote{Here, all forms are written with bold letters,
$\mathbf A = \frac{1}{p!}A_{\mu^1\cdots \mu^p}dx^{\mu^1}\wedge
\dots \wedge dx^{\mu^p}$. The convention for the Hodge dual of a
$p$-form $\omega^p$ is $\star\, \omega^p =
\sqrt{|g|}\,\omega^{\mu_1\dots \mu_p}(d^{n-p}x)_{\mu_1\dots\mu_p}$
with $(d^{n-p}x)_{\mu_1\dots\mu_p} \hat = \frac 1{p!(n-p)!}\,
\epsilon_{\mu_1\dots \mu_p \mu_{p+1}\cdots \mu_n}
dx^{\mu_{p+1}}\dots dx^{\mu_n}$. Hence, if $\alpha^{(p)}$ and
$\beta^{(q)}$ are $p$ and $q$ forms with $q \leq p \leq n$, one
has $\beta^{(q)} \wedge \star\, \alpha^{(p)} =
\frac{1}{q!}\alpha^{(p)\mu_1\cdots \mu_{p-q}\rho_1 \cdots
\rho_q}\beta^{(q)}_{\rho_1 \cdots
\rho_q}(d^{n-(p-q)}x)_{\mu_1\cdots \mu_{p-q}}$. The inner product
$i_{\xi}\omega^{(n-1)}$ can be written explicitly as $( \xi^\nu
\omega^{(n-1)\mu} - \xi^\mu
\omega^{(n-1)\nu})(d^{n-2}x)_{\mu\nu}$.}. The fields of the theory
are collectively denoted by $\phi^i \equiv (g_{\mu\nu}, \mathbf
A,\chi)$. We will set $16\pi G = 1$ for convenience.

\section{Conservation laws}

A very convenient mathematical setting to handle with $n-1$ or
$n-2$-form conservation laws or more generally $(n-q)$-form
conservation laws ($0 \leq q < n $) is the study of local
cohomology in field theories~\cite{Andersonbook,Barnich:2000zw},
see~also~\cite{Torre:1997cd} for an introduction. A conservation
law consists in the existence of a $(n-q)$-form $k^{(n-q)}$ which
is conserved on-shell $d k^{(n-q)} \approx 0$ and which is
non-trivial, i.e. not the differential of another form on-shell,
$k^{(n-q)} \napprox d (\cdot)$.

In Minkowski spacetime $g_{\mu\nu}= \eta_{\mu\nu}$, $\chi = 0$ and
for a trivial bundle $\mathbf A$, all these lower degree conserved
forms are classified by the characteristic cohomology of $p$-form
gauge theories~\cite{Henneaux:1996ws}. These laws are generated in
the exterior product by the forms $\star \mathbf H$ dual to the
field strength~\footnote{When magnetic charges are allowed, there
are additional conserved quantities as $\oint \mathbf H \neq 0$.
However, the field strength $\mathbf H$ cannot be written as the
derivative of a potential $\mathbf B$ and the action principle has
to be modified.
This case will not be treated below.}. More precisely, for odd
$n-p-1$, one can construct the conserved $n-p-1$-form $\star
\mathbf H$. For even $n-p-1$, factors $\star \mathbf H$ mutually
commute and one may construct the conserved  forms $l(n-p-1)$
$\underbrace{\star \mathbf H \wedge \dots \wedge \star\mathbf
H}_l$ for any integer $l$ such that $l(n-p-1) < n-1$.

When gravity and the scalar field are present, the charges
\begin{eqnarray}
\mathbf Q^{(n-p-1)} &= e^{-\alpha \chi} \star \mathbf H, \qquad
\qquad n-p-1\;\text{odd}\label{elc_ch}
\\\label{ch_other}\mathbf  Q^{l(n-p-1)}& = e^{-l\alpha \chi}
\underbrace{\star \mathbf H \wedge \dots \wedge \star\mathbf
H}_l,\qquad n-p-1\;\text{even}
\end{eqnarray}
still enumerate the non-trivial conservation
laws~\cite{Barnich:1995ap,Henneaux:1996ws} \footnote{The
conservations laws that we consider here are called dynamical
because they explicitly involve the equations of motion. It exists
also specific topological conservation laws, see
e.g.~\cite{Torre:1994pf}.}

In order to investigate the first law of thermodynamics, where
variations around a solution are involved, we now extend the
analysis to the linearized theory.

In linearized gravity, only $(n-2)$-form conservation laws are
allowed~\cite{Barnich:1995db}. The classification of non-trivial
conserved $(n-2)$-forms was described in \cite{Barnich:2001jy} and
is straightforward to specialize in our case. The equivalence
classes of conserved $(n-2)$-forms of the linearized theory for
the variables $\delta\phi^i$ around a fixed reference solution
$\phi^i$ are in correspondence with equivalence classes of gauge
parameters $\xi^\mu(x),\mathbf \Lambda(x)$ satisfying the
reducibility equations $\delta_{\xi,\mathbf\Lambda} \phi^i =
0$~\footnote{This correspondence is one-to-one for gauge
parameters that may depend on the linearized fields $\varphi^i$
and that
satisfy~$\delta_{\xi(x,\varphi^i),\mathbf\Lambda(x,\varphi^i)}
\phi^i \approx_{lin} 0$, i.e. zero for solutions $\varphi^i$ of
the linearized equations of motion. However, it has been proven
in~\cite{Barnich:2004ts} that this $\varphi$-dependence is not
relevant in the case of Einstein gravity. Such a dependence will
not be considered here.}, i.e.
\begin{eqnarray}
\left\{\begin{array}{c}
  \cL_\xi  g_{\mu\nu} = 0,\\ \cL_\xi  {\mathbf A} + d
 \mathbf \Lambda = 0,\\
  \cL_\xi  \chi = 0.
\end{array}\right. \label{eq:red}
\end{eqnarray}

In this paper, we construct a $(n-2)$-form $\mathbf k_{\xi,\mathbf
\Lambda}$ enjoying the following properties. First, for each
generalized Killing vector $(\xi,\mathbf \Lambda)$ satisfying the
reducibility equations~\eqref{eq:red}, the surface form $\mathbf
k_{\xi,\mathbf \Lambda}$ will be closed on-shell. As a result, the
infinitesimal charge difference between solutions $\phi^i$ and
$\phi^i+\delta \phi^i$ associated with any parameter $(
\xi,\mathbf \Lambda)$ satisfying~\eqref{eq:red},
\begin{eqnarray}
  \delta \cQ_{ \xi, \mathbf\Lambda} \hat =
  \oint_S \mathbf k_{ \xi,\mathbf \Lambda}[ \delta \phi ;
  \phi],\label{ch_to}
\end{eqnarray}
will only depend on the homology class of $S$. Second, since the
$(n-2)$-form will be build from the weakly vanishing Noether
current, the usual ambiguities that should be treated with care in
covariant phase space methods~\cite{Iyer:1994ys} will be avoided
here\footnote{Indeed, by construction, the weakly vanishing
Noether current does not depend on boundary terms that may be
added to the Lagrangian. Moreover, if one adds a weakly vanishing
exact $(n-1)$-form $d l^{(n-2)}$ to the Noether current, the
resulting $(n-2)$-form will be supplemented by an irrelevant exact
term and by $\delta l^{(n-2)}$ which vanishes on-shell.}. For
additional properties of these surface charges, as the
representation theorem of the Lie algebra of reducibility
parameters, the reader is referred to the original
work~\cite{Barnich:2001jy,Barnich:2003xg}.

\section{Surface forms}

Following the lines of \cite{Barnich:2001jy,Barnich:2003xg}, one
can construct the weakly vanishing Noether currents associated
with the couple $(\xi,\mathbf \Lambda)$ by integrating by parts
the expression~$\delta_{\xi,\mathbf\Lambda} \phi^i \varQ{\mathbf
L}{\phi^i}$ and using the Noether identities. We obtain
\begin{eqnarray}
  \label{eq:4bis}
\mathbf S_{\xi,\mathbf\Lambda} &=& \star \bigg(
(-2G_\mu^{\;\,\nu}+ T_{\mathbf A\,\mu}^{\,\;\,\nu}+
T_{\chi\;\mu}^{\;\,\nu} )\xi_\nu dx^\mu \label{eq:S1}
\\
&-&  \frac{1}{(p-1)! }D_\beta(e^{-\alpha
\chi}H_\mu^{\;\,\beta\mu^1\cdots \mu^{p-1}})(\xi^\rho A_{\rho\mu^1
\cdots \mu^{p-1}}+\Lambda_{\mu^1 \cdots \mu^{p-1}})dx^\mu
\bigg),\nonumber
 \end{eqnarray}
where the stress tensors are given by
\begin{eqnarray}
T_{\mathbf A}^{\mu\nu} &=& e^{-\alpha \chi} \left( \frac{1}{p!}
H^{\mu}_{\;\;\mu^1 \cdots \mu^{p}}H^{\nu\mu^1 \cdots \mu^{p}} -
\frac{1}{2(p+1)!}g^{\mu\nu}H^2 \right) ,\\
T_\chi^{\mu\nu} &=& (\d^\mu \chi \d^\nu \chi -
\frac{1}{2}g^{\mu\nu}\d^\alpha \chi \d_\alpha\chi).
\end{eqnarray}
The surface form~$\mathbf k_{\xi,\mathbf\Lambda}[\delta
\phi;\phi]=k_{\xi,\mathbf\Lambda}^{[\mu\nu]} (d^{n-2}x)_{\mu\nu}$
can be obtained as a result of a contracting homotopy $\mathbf
I^{n-1}_{\delta \phi}$ acting on the current $\mathbf
S_{\xi,\mathbf\Lambda}$, see
e.g.~\cite{Andersonbook,Barnich:2003xg}. Using the following
property of the homotopy operators,
\begin{equation}
d \mathbf I_{\delta \phi}^{q-1}\mathbf \omega^{(q-1)}+ \mathbf
I_{\delta \phi}^{q} d\mathbf \omega^{(q-1)}= \delta\mathbf
\omega^{(q-1)},\qquad \forall \mathbf \omega^{(q-1)},\quad q \leq
n ,\label{prop_I}
\end{equation}
one has
\begin{equation}
d \mathbf k_{\xi,\mathbf\Lambda} = \delta \mathbf
S_{\xi,\mathbf\Lambda} - \mathbf I_{\delta
\phi}^{n-2}\left(\delta_{\xi,\mathbf\Lambda} \phi^i \varQ{\mathbf
L}{\phi^i} \right).
\end{equation}
The closure $d \mathbf k_{\xi,\mathbf\Lambda}[\delta \phi;\phi]
\approx 0$ then hold whenever $\phi^i$ satisfies the equations of
motion, $\delta \phi^i$ the linearized equations of motion and
$(\xi,\mathbf\Lambda)$ the system \eqref{eq:red}.

Let us now split the current into different contributions,
$\mathbf S_{\xi,\mathbf\Lambda} =\mathbf S^{g}_\xi + \mathbf
S^{\chi}_{\xi} + \mathbf S^{\mathbf A}_{\xi,\mathbf\Lambda}$ with
\begin{eqnarray}
\mathbf S^{g}_\xi &=&  \star (-2G_\mu^{\;\,\nu}\xi_\nu \,dx^\mu), \\
\mathbf S^{\chi}_{\xi} &=&\star ( T_{\chi \; \mu}^{\;\,\nu}
\xi_\nu \, dx^\mu),
\end{eqnarray}
and $\mathbf S^{\mathbf A}_{\xi,\mathbf\Lambda}$ being the
remaining expression. Since the homotopy $\mathbf I^{n-1}_{\delta
\phi}$ is linear in its argument, the surface form can be
decomposed as $\mathbf k_{\xi, \mathbf\Lambda} = \mathbf
k^{g}_{\xi}+ \mathbf k^{\chi}_{\xi} + \mathbf k^{\mathbf A}_{\xi,
\mathbf\Lambda}$.

The gravitational contribution $\mathbf k^{g}_{\xi}$, which
depends only on the metric and its deviations, coincides with the
Abbott-Deser expression \cite{Abbott:1982ff} and, for Killing
vectors, with the expression derived in the Hamiltonian approach
of Regge-Teitelboim~\cite{Regge:1974zd}. It can be written as
\begin{eqnarray}
\mathbf k^{g}_{ \xi}[\delta g;g] &=& -\delta \mathbf Q^g_{ \xi}+
\mathbf Q^g_{\delta \xi} -i_{\xi}\mathbf \Theta[\delta g] -\mathbf
E_\cL[\cL_\xi g, \delta g],\label{grav_contrib}
\end{eqnarray}
where
\begin{eqnarray}
\mathbf Q^g_{\xi}&=& \star \Big( \half (D_\mu\xi_\nu-D_\nu\xi_\mu)
dx^\mu \wedge dx^\nu \Big),\label{Komar_term}
\end{eqnarray}
is the Komar $n-2$ form and
\begin{eqnarray}
\mathbf \Theta[\delta g]&=&\star \Big(  (D^\sigma \delta
g_{\mu\sigma}-
g^{\alpha\beta} D_\mu \delta g_{\alpha\beta})\,dx^\mu\Big),\\
\mathbf E_\cL[\cL_\xi g, \delta g] &=& \star \Big( \half \delta
g_{\mu\alpha}(D^\alpha \xi_\nu + D_\nu \xi^\alpha) dx^\mu \wedge
dx^\nu \big).
\end{eqnarray}
The supplementary term, $E_\cL$, with respect to the Iyer-Wald
form~\cite{Iyer:1994ys} vanishes for Killing vectors.

The scalar contribution is easily found to be $\mathbf k^{\chi}_{
\xi}[\delta g,\delta\chi;g,\chi] = i_\xi \mathbf
\Theta_\chi$~\cite{Barnich:2002pi} with
\begin{equation}
\mathbf \Theta_\chi = \star ( d\chi \, \delta \chi )
.\label{phicharge}
\end{equation}

Let us now compute the contribution $\mathbf k^{\mathbf A}_{\xi,
\mathbf\Lambda}$ from the $p$-form. After some algebra, one can
rewrite the current $\mathbf S^{\mathbf A}_{\xi,\mathbf\Lambda}$
as
\begin{eqnarray}
\mathbf  S^{\mathbf A }_{\xi,\mathbf\Lambda}&=&- d \mathbf
Q^{\mathbf A}_{\xi,\mathbf \Lambda} + e^{-\alpha \chi}(\cL_\xi
\mathbf A+d\mathbf \Lambda )\wedge \star \mathbf H - \half
e^{-\alpha \chi } i_\xi(\mathbf H \wedge \star \mathbf H)
\end{eqnarray}
with
\begin{equation}
\mathbf Q^{\mathbf A}_{ \xi,\mathbf\Lambda}= e^{-\alpha \chi} (
i_\xi \mathbf A +\mathbf \Lambda ) \wedge \star\mathbf
H.\label{def_QA}
\end{equation}
Using the property~\eqref{prop_I}, the surface form $\mathbf
k^{\mathbf A}_{ \xi,\mathbf\Lambda}$ reduces to
\begin{eqnarray}
\mathbf k^{\mathbf A}_{ \xi,\mathbf\Lambda} &=& -\delta\mathbf
Q^{\mathbf A}_{\xi,\mathbf\Lambda}  + \mathbf Q^{\mathbf
A}_{\delta\xi,\delta\mathbf\Lambda}+ d \mathbf
I^{n-2}_{\delta\phi}\mathbf Q^{\mathbf A}_{\xi,\mathbf\Lambda} \nonumber\\
&& + \mathbf I_{\delta \phi}^{n-1}\big( e^{-\alpha \chi}(\cL_\xi
\mathbf A+d\mathbf \Lambda )\wedge \star \mathbf H - \half
e^{-\alpha \chi } i_\xi(\mathbf H \wedge \star \mathbf H) \big),
\end{eqnarray}
where the exact term $d \mathbf I^{n-2}_{\delta\phi}\mathbf
Q^{\mathbf A}_{\xi,\mathbf\Lambda}$ is trivial and can be dropped.
The last term can then be computed easily since it admits only
first derivatives of the gauge potential. The homotopy thus
reduces in that case to $I^{n-1}_{\delta \mathbf A} = \half \delta
\mathbf A \frac{\partial}{\partial \mathbf H}$. We eventually get
\begin{equation}
\mathbf k^{\mathbf A}_{ \xi,\mathbf\Lambda}[\delta g,\delta
\mathbf A,\delta \chi ;g,\mathbf A,\chi ]=-\delta \mathbf
Q^{\mathbf A}_{\xi,\mathbf\Lambda} + \mathbf Q^{\mathbf
A}_{\delta\xi,\delta\mathbf\Lambda} + i_\xi \mathbf
\Theta_{\mathbf A}-\mathbf E^{\mathbf A}_\cL[\cL_\xi \mathbf A+d
\mathbf\Lambda,\delta \mathbf A] \label{Bcharge}
\end{equation}
with
\begin{eqnarray}
\mathbf \Theta^{\mathbf A} &=& e^{-\alpha \chi} \delta \mathbf A
\wedge \star \mathbf H,\label{ThetaA}\\
\mathbf E^{\mathbf A}_\cL[\cL_\xi \mathbf A+d
\mathbf\Lambda,\delta \mathbf A] &= & e^{-\alpha \chi} \star \big(
\half \frac{1}{(p-1)!}\delta
\mathbf A_{\mu\alpha_1\cdots \alpha_{p-1}} \nonumber \\
&&\hspace{-30pt}(\cL_\xi \mathbf A+d\mathbf
\Lambda)_\nu^{\;\,\,\alpha_1\cdots \alpha_{p-1}} dx^\mu\wedge
dx^\nu \big)
\end{eqnarray}
which has a very similar structure as the gravitational field
contribution~\eqref{grav_contrib}. For reducibility
parameters~\eqref{eq:red}, the term involving $\cL_{\xi} \mathbf
A+d \mathbf \Lambda$ vanishes. The form~\eqref{def_QA} will be
referred to as a Komar term, in analogy with the gravitational
Komar term~\eqref{Komar_term}.

For $p=1$ and reducibility parameters, the surface
form~\eqref{Bcharge} reduces to the well-known expression for
electromagnetism, see e.g. \cite{Gao:2001ut}.
Expression~\eqref{Bcharge} and the one derived
in~\cite{Rogatko:2005aj,Rogatko:2006xh} have a similar structure
but differ in two respects. First, our surface form contains the
additional term $\mathbf E^{\mathbf A}_\cL[\cL_\xi \mathbf A+d
\mathbf\Lambda,\delta \mathbf A]$. Nevertheless, since this term
vanishes for reducibility parameters, it will not be relevant for
exact conservation laws. Second, the form factors in the Komar
term $\mathbf Q^{\mathbf A}_{ \xi,\mathbf\Lambda}$ differ
from~\cite{Rogatko:2005aj,Rogatko:2006xh}. The results
of~\cite{Rogatko:2005aj,Rogatko:2006xh} agree with ours when the
right-hand side of equation~(10) of~\cite{Rogatko:2005aj} and
equation (4) of~\cite{Rogatko:2006xh} are multiplied by
$-\frac{p+1}{2}$.

Let us assume that~\eqref{eq:red} holds for a field configuration
$(g,\mathbf A,\chi)$. As a consistency check, note that the
surface form~\eqref{Bcharge} satisfies the equality on-shell
$\mathbf k^{\mathbf A}_{ \xi,\mathbf\Lambda}[\delta g=0,\delta
\mathbf A =d \omega^{(p-1)},\delta \chi=0;g,\mathbf A,\chi]
\approx d(\cdot)$. The charge difference~\eqref{ch_to} between two
configurations differing by a gauge transformation $\delta \mathbf
A= d \omega^{p-1}$, is thus zero on-shell.

Besides generalized Killing vectors $(\xi,\mathbf \Lambda)$ which
are also symmetries of the gauge field and of the scalar $\chi$,
there may be charges associated with non-trivial gauge parameters
$(\xi=0,\mathbf \Lambda \neq d(\cdot))$. For $p=1$, in
electromagnetism, $\mathbf \Lambda = constant \neq 0$ is such a
parameter and the associated charge is the electric
charge~\eqref{elc_ch}. For $p>1$, non-exact forms $\mathbf
\Lambda$ may exist if the topology of the manifold is non-trivial.
The charges with a non-trivial closed form $\mathbf \Lambda$ which
does not vary along solutions is given by
\begin{equation}
\cQ_{0,-\mathbf \Lambda} = \oint_S e^{-\alpha \chi} \mathbf
\Lambda \wedge \star \mathbf H = \oint_{T} e^{-\alpha \chi} \star
\mathbf H,\label{dipole_ch}
\end{equation}
where $S$ is a $n-2$ surface enclosing the non-trivial cycle $T$
dual to the form $\mathbf \Lambda$. It is simply the integral
of~\eqref{elc_ch} on the non-trivial cycle. The
charges~\eqref{dipole_ch} are thus the generalization for
$p$-forms of electric charges.

The properties of the surface form~\eqref{Bcharge} under
transformations of the potential $\mathbf A$ are worth mentioning.
The transformation $\mathbf A \rightarrow \mathbf A + d\epsilon$
preserves the reducibility equations~\eqref{eq:red} if $d\cL_\xi
\mathbf\epsilon = 0$. In that case, $\cL_\xi\mathbf \epsilon$ can
be written as the sum of an exact form and an harmonic form that
we denote as $f(\mathbf\epsilon,\xi) \mathbf \Lambda^\prime$ with
$\mathbf \Lambda^\prime$ not varying along solutions, $\delta
\mathbf \Lambda^\prime= 0$ and $f(\mathbf\epsilon,\xi)$ constant.
In Einstein-Maxwell theory, one has $\mathbf \Lambda^\prime = 1$
and $f(\mathbf\epsilon,\xi)=\cL_\xi\mathbf \epsilon$. Under the
transformation $\mathbf A \rightarrow \mathbf A + d\epsilon$, the
surface form~\eqref{Bcharge} changes according to
\begin{equation}
\mathbf k^{\mathbf A}_{ \xi,\mathbf\Lambda} \rightarrow \mathbf
k^{\mathbf A}_{ \xi,\mathbf\Lambda} -f(\epsilon,\xi)\mathbf
\delta(\mathbf \Lambda^\prime \wedge e^{-\alpha \chi}\star \mathbf
H) + d(\cdot)+\mathbf t_{\xi},\qquad \mathbf t_{\xi} \approx 0.
\end{equation}
Defining the charge associated to $\mathbf \Lambda^\prime$
as~\eqref{dipole_ch}, one sees that the infinitesimal
charge~\eqref{ch_to} varies on-shell as
\begin{equation}
\delta\cQ_{ \xi,\mathbf\Lambda} \rightarrow \delta\cQ_{
\xi,\mathbf\Lambda} - f(\mathbf\epsilon,\xi) \delta\cQ_{0,-\mathbf
\Lambda^\prime}.\label{non-gauge_inv}
\end{equation}
As a consequence, a transformation $\mathbf A \rightarrow \mathbf
A + d\epsilon$ admitting a non-vanishing function
$f(\mathbf\epsilon,\xi)$ cannot be considered as a gauge
transformation because such a transformation does not leave the
conserved charges of the solution invariant.

\section{First law}

We now assume that $\phi^i$ and $\phi^i+\delta \phi^i$ are
stationary black hole solutions with Killing horizon. The
generator of the Killing horizon of $\phi^i$, $\xi = \d_t +
\Omega^a \d_{\varphi_a}$ is a combination of the Killing vectors
$\d_t$ and $\d_{\varphi_a}$, $a=1 \dots \lfloor (n-1)/2 \rfloor$.
The variation of energy $\delta \cE$ and angular momenta $\delta
\cJ_a$ are defined as the charges associated with the Killing
vectors $\d_t$ and $-\d_{\varphi_a}$, respectively~\footnote{The
relative sign difference between the definitions of $\delta
\mathcal E$ and $\delta \mathcal J^a$ trace its origin to the
Lorentz signature of the metric \cite{Iyer:1994ys}.}. Remark that
this definition of energy is more natural than the one used in
\cite{Rogatko:2005aj,Rogatko:2006xh}, where a factor $\alpha =
\frac{n-3}{n-2}$ was artificially added in equation (16) of
\cite{Rogatko:2005aj} and in equation (8) of
\cite{Rogatko:2006xh}.

We assume that $\xi$ is a solution of~\eqref{eq:red} with $\mathbf
\Lambda = 0$. We also require that $\xi + \delta\xi$ is a symmetry
of the perturbed black hole $\phi^i+\delta\phi^i$.

The first law is then a consequence of the equality~\footnote{The
first law can be straightforwardly generalized to reducibility
parameters satisfying $\cL_\xi \mathbf A + \mathbf \Lambda = 0$
with $\mathbf\Lambda \neq d(\cdot)$. This simply amounts to add a
contribution at infinity and at the horizon.}
\begin{eqnarray}
  \oint_{S^\infty}\mathbf  k_{ \xi,0}[ \delta \phi ; \phi]
  =\oint_{H}\mathbf  k_{ \xi,0}[ \delta \phi; \phi],
\end{eqnarray}
where $S^\infty$ is a $(n-2)$-sphere at infinity and $H$ is any
cross-section of the Killing horizon.

Using the linearity of $\mathbf k_{ \xi,0}$ with respect to $\xi$,
the left-hand side is simply given by $\delta \cE - \Omega^a
\delta \cJ_a$. Splitting the right-hand side, we get
\begin{eqnarray}
\delta \cE - \Omega^a \delta \cJ_a  =\oint_{H}\mathbf  k^{g}_{
\xi,0}[ \delta \phi; \phi]+\oint_{H}\mathbf  k^{\chi}_{ \xi,0}[
\delta \phi; \phi]+\oint_{H}\mathbf  k^{\mathbf A}_{ \xi,0}[
\delta \phi; \phi].\label{Threet}
\end{eqnarray}

The geometric properties of the Killing horizon then allow one to
express the pure gravitational contribution into the
form~\cite{Bardeen:1973gs,Wald:1993nt,Iyer:1994ys,Jacobson:1993vj}
\begin{eqnarray}
\oint_{H}\mathbf  k^{g}_{ \xi,0}[ \delta \phi; \phi]=
\frac{\kappa}{8\pi G} \delta \cA,\label{eq:geo_t}
\end{eqnarray}
where $\kappa$ is the surface gravity and $\cA$ the area of the
black hole and where $G$ factors have been restored. Here, the
cross-section of the horizon could be chosen to lie on the future
horizon or, when it exists, to be the bifurcation surface $H_B$.
See also~\cite{Compere:2006my} for a derivation of the first
law~\eqref{eq:geo_t} for stationary perturbations on the future
event horizon without assumption on the way to perform the
variation.

It is now convenient for the rest of the computation to choose a
cross-section lying on the future horizon. The integration measure
for the $(n-2)$-forms then becomes
\begin{equation}
\sqrt{|g|}(d^{n-2}x)_{\mu\nu} = \frac{1}{2} (\xi_\mu n_\nu - n_\mu
\xi_\nu) d\mathcal{A},\label{mesure}
\end{equation}
where $d\mathcal{A}$ is the angular measure and $n^\mu$ is an
arbitrary null vector transverse to the horizon normalized with
$n^\mu \xi_\mu = -1$, see e.g. equations~(6.14) and (6.70)
of~\cite{Townsend:1997ku} for details.

Using~\eqref{phicharge}, the scalar contribution can be written as
\begin{equation}
\oint_{H}\mathbf  k^{\chi}_{ \xi,0}[ \delta \phi; \phi] = -
\oint_H d\cA\, \delta \chi (\cL_\xi \chi + \xi^2 \cL_n \chi) =
0,\label{contrib_chi2}
\end{equation}
which vanishes thanks to the reducibility
equations~\eqref{eq:red}, assuming the regularity of the scalar
field on the horizon. By continuity, this result is also valid on
the bifurcation surface $H_B$.

The contribution of the $p$-form can be computed using the
arguments of~\cite{Gauntlett:1998fz,Copsey:2005se}. The
Raychaudhuri equation gives $R_{\mu\nu}\xi^\mu \xi^\nu = 0$ on the
horizon. It follows by Einstein's equations and by the identity
$\cL_\xi \phi = 0$ that $i_\xi \mathbf{H}$ has vanishing norm on
the horizon. But as $i_\xi (i_\xi\mathbf H) = 0$, $i_\xi \mathbf
H$ is tangent to the horizon. $i_\xi\mathbf H$ has thus the form
$\mathbf\xi \wedge \dots \wedge \mathbf\xi$ by antisymmetry of
$\mathbf H$ and its pullback to the horizon vanishes. The equation
$\cL_\xi \mathbf A = 0$ can be written as $d i_\xi \mathbf A  =
-i_\xi \mathbf H$. Therefore, the pull-back of $i_\xi  \mathbf A$
on the horizon is a closed form.

For $p=1$, $-i_\xi \mathbf A = \Phi$ is simply the scalar electric
potential at the horizon. When $p > 1$, the quantity $-i_\xi
\mathbf A $ pulled-back on the horizon is the sum of an exact form
$d \mathbf{e}$ and an harmonic form $\mathbf h$. If the horizon
has non-trivial $n-p-1$ cycles $T_a$, one can define the harmonic
forms dual to $T_a$ by duality between homology and cohomology as
\begin{equation}
\int_{T_a} \mathbf\sigma = \int_H \mathbf\Omega_a \wedge
\mathbf\sigma , \qquad \forall \mathbf\sigma.
\end{equation}
The harmonic form $\mathbf h$ is then a sum of terms $\mathbf h =
\Phi^a \mathbf\Omega_a$ with $\Phi^a$ constant over the
non-trivial cycles.

The contribution from the potential contains three
terms~\eqref{Bcharge}. The Komar term~\eqref{def_QA} can be
written as
\begin{eqnarray}
\oint_H \mathbf Q^{\mathbf A}_{\xi,0} = -\Phi^a \oint_{T_a}
e^{-\alpha \chi}\star \mathbf H,
\end{eqnarray}
where the exact form $d \mathbf{e}$ do not contribute on-shell.
 We recognize on the right-hand side the
conserved form written in~\eqref{dipole_ch}. Let us denote by
$\cQ_a$ the integral $\oint_{T_a} e^{-\alpha \chi}\star \mathbf
H$.

Using~\eqref{mesure}, the contribution~$\oint_H i_\xi \mathbf
\Theta_{\mathbf A}[\delta \phi,\phi]$ reads as
\begin{equation}
\oint_H i_\xi \mathbf \Theta_{\mathbf A}[\delta \phi,\phi] =
\oint_H e^{-\alpha\chi}(i_\xi \delta \mathbf A) \wedge \star
\mathbf H -\oint_H d\cA\, \xi^2 \star \Big( \delta \mathbf A
\wedge \star (i_n \mathbf H) \Big).\label{eq:k10}
\end{equation}
The first term of~\eqref{eq:k10} nicely combines with the second
term of~\eqref{Bcharge} into $-\oint_{T_a} \delta \Phi^a
e^{-\alpha \chi}\star \mathbf H = -\delta \Phi^a \cQ_a$ because
$\delta \Phi^a$ is constant as a consequence of the hypotheses on
the variation. In the second term of~\eqref{eq:k10}, one can
replace $\delta \mathbf A$ by its pull-back $\phi_* \delta \mathbf
A$ on the future horizon. Indeed, decomposing $\delta\mathbf A =
\mathbf n \wedge \omega^{(1)}+\phi_*\delta\mathbf A$, one sees
that the term involving $\mathbf n$ do not contribute because of
the antisymmetry of $\mathbf H$. Therefore, the second term
in~\eqref{eq:k10} will vanish if $\mathbf H$ is regular and if the
pull-back $\phi_* \delta \mathbf A$ on the future horizon is
regular.

Finally, the contribution from the potential on the horizon
reduces to
\begin{equation}
\oint_{H}\mathbf  k^{\mathbf A}_{ \xi,0}[ \delta \phi; \phi] =
\Phi^a \delta \cQ_a,\label{contrib_ch}
\end{equation}
as it should to give the first law
\begin{equation}
\delta \cE - \Omega^a \delta \cJ_a= \frac{\kappa}{8\pi G}\delta
\cA + \Phi^a \delta \cQ_a.\label{first_law_3}
\end{equation}
Since the computation can be done entirely on the future horizon,
this first law is valid in the extremal case, with $\kappa =0$.
The relations~\eqref{eq:geo_t} and~\eqref{contrib_chi2} hold on
any cross-section of the horizon. Since the surface
charges~\eqref{ch_to} only depend on the homology class of the
surface $S$, the third term in the right-hand side
of~\eqref{Threet} has to be equal to~\eqref{contrib_ch} for any
cross-section of the horizon as well. Therefore, when the
bifurcation surface exists and when the regularity hypotheses are
fulfilled, the first law~\eqref{first_law_3} also holds there.

\section{Application to black rings}

Let us consider the black ring with dipole charge described
in~\cite{Emparan:2004wy}. This black ring is a solution to the
action~\eqref{act1} in five dimensions for a two-form $\mathbf A$.
The solution admits three independent parameters: the mass, the
angular momentum and a dipole charge $\oint_{S^2} e^{-\alpha \chi}
\star \mathbf H$ where $S^2$ is a two-sphere section of the black
ring whose topology is $S^2 \times S^1$.

The thermodynamics of this solution was worked out in the original
paper~\cite{Emparan:2004wy}. The role of dipole charges in the
formalism of Sudarsky and Wald \cite{Sudarsky:1992ty} was
elucidated in~\cite{Copsey:2005se}. The metric, the scalar field
and the gauge potential are written in equations~(3.2)-(3.3)-(3.4)
of~\cite{Copsey:2005se}. There, the gauge potential
\begin{equation}
\mathbf A = B_{t\psi} dt \wedge d\psi,
\end{equation}
was shown to be singular on the bifurcation surface in order to
avoid a delta function in the field strength on the black ring
axis. Here, we point out that this singularity in the potential
does not prevent one from studying thermodynamics on the future
event horizon along the lines above since the pull-back of the
potential is regular there.

Indeed, following~\cite{Emparan:2006mm}, one can introduce ingoing
Eddington-Finkelstein coordinates near the horizon of the black
ring as
\begin{eqnarray}
d\psi &=& d\psi^\prime + \frac{dy}{G(y)}\sqrt{-F(y)H^N(y)},\\
dt &=& dv - C D R \frac{(1+y)\sqrt{-F(y)H^N(y)} }{F(y) G(y)}dy.
\end{eqnarray}
The metric is regular in these coordinates and the gauge potential
can be written as
\begin{equation}
\mathbf A = B_{t\psi}dv \wedge d\psi^\prime + dy \wedge
\omega^{(1)},
\end{equation}
for some $\omega^{(1)}$. The pull-back of the gauge potential to
the future horizon $y = -1/\nu$ is explicitly regular because
$B_{t\psi}$ is finite and $v$ and $\psi^\prime$ are good
coordinates.

The first law for black rings may then be seen as a consequence
of~\eqref{first_law_3}.

\section{Application to black strings in plane waves}

We now turn to the definition of mass in asymptotic plane wave
geometries. Here, we show that the integration of the surface form
$\mathbf k_{\d_t,0}[\delta \phi,\phi]$ along a path $\gamma$ in
solution space~\cite{Wald:1999wa,Barnich:2003xg},
\begin{equation}
\cE = \int_\gamma \oint_{S^\infty} \mathbf k_{\d_t,0}[\delta
\phi,\phi]
\end{equation}
provides a natural definition of mass, satisfying the first law of
thermodynamics.

The action of the NS-NS sector of bosonic supergravity in
$n$-dimensions in string frame reads
\begin{eqnarray} S[G,B,\phi_s] &=&  \frac{1}{16\pi G} \int d^{n}
x \sqrt{-G} e^{-2\phi_s} \left[ R_G +4\d_\mu\phi_s \d^\mu \phi_s
\nonumber - \frac{1}{12}H^2 \right],
\end{eqnarray}
when all fields in the $D-n$ compactified dimensions vanish. In
Einstein frame, $g_{\mu\nu} = e^{-4\tilde\phi/(n-2)} G_{\mu\nu}$,
$\phi = \alpha \phi_s$, the action can be written as~\eqref{act1}
with $\alpha = \sqrt{8/(n-2)}$ and $\mathbf A = B$.

Neutral black string in the $n$-dimensional maximally symmetric
plane wave background $\cP_n$, with $n> 4$, are given by
\cite{Gimon:2003xk,Hubeny:2003ug,Hashimoto:2004ve}
\begin{eqnarray}
ds_s^2 &=& -\frac{f_n(r)(1+\beta^2
r^2)}{k_n(r)}dt^2-\frac{2\beta^2 r^2 f_n(r)}{k_n(r)}dt dy +r^2
d\Omega^2_{n-3} \nonumber\\\nonumber &+& \left( 1-\frac{\beta^2
r^2}{k_n(r)}\right) dy^2 + \frac{dr^2}{f_n(r)}-\frac{r^4
\beta^2(1-f_n(r))}{4k_n(r)}\sigma^2_n,\\
e^{\phi_s} &=& \frac{1}{\sqrt{k_n(r)}}, \qquad B= \frac{\beta
r^2}{2k_n(r)}(f_n(r)dt+dy)\wedge \sigma_n
\end{eqnarray}
where
\begin{equation}
f_n(r) = 1-\frac{M}{r^{n-4}}, \qquad k_n(r) = 1+\frac{\beta^2
M}{r^{n-6}}.
\end{equation}
The black strings have horizon area per unit length given by $\cA
= M^{\frac{n-3}{n-4}}A_{n-3}$ where
\begin{equation}
A_{n-3} = \frac{2\pi^{\frac{n-2}{2}}}
{\Gamma\left(\frac{n-2}{2}\right)},
\end{equation}
is the area of the $n-3$ sphere. Choosing the normalization of the
horizon generator as $\xi = \d_t$, the surface gravity is given by
$\kappa = \sqrt{-1/2 (D_\mu \xi_\mu D^\mu \xi^\nu)} =
\frac{n-4}{2}M^{-\frac{1}{n-4}}$.

Using the surface forms defined above, the charge difference
 associated with $\Q{}{t}$ between two infinitesimally close black string solutions $\phi$,
$\phi+\delta \phi$ is given by
\begin{equation}
\delta \cQ_{\d_t} = \oint k_{\d_t,0}[\delta \phi,\phi] =
\frac{n-3}{16\pi G } A_{n-3} \delta M,
\end{equation}
which reproduces the expectations of
\cite{Gimon:2003xk,Hubeny:2003ug,Hashimoto:2004ve}. This quantity
is integrable and allows one to define $\cQ_{\d_t} =
\frac{n-3}{16\pi G } A_{n-3} M$ where the normalization of the
background has been set to zero. It is easy to check that the
first law is satisfied.

Note that one freely can choose a different normalization for the
generator $\xi^\prime = N \d_t$. In that case, the surface gravity
changes according to $\kappa^\prime = N \kappa$, the charge
associated to $\xi^\prime$ becomes $\delta \cQ_{\xi^\prime} =
\frac{n-3}{16\pi G } A_{n-3} \,N\,\delta M$ and the first law is
also satisfied. However, $N$ cannot be a function of $\beta$.
Otherwise, the charge $\cQ_{\xi^\prime}$ would not be defined.

\section*{Acknowledgments} The author thanks G.~Barnich, K.~Copsey and
M. Henneaux for their valuable comments. The author is Research
Fellow at the National Fund for Scientific Research (FNRS
Belgium). This work is also supported in part by a ``P{\^o}le
d'Attraction Interuniversitaire'' (Belgium), by IISN-Belgium,
convention 4.4505.86, by Proyectos FONDECYT 1970151 and 7960001
(Chile) and by the European Commission program
MRTN-CT-2004-005104, in which the author is associated to
V.U.~Brussel.


\end{document}